\newcommand{\cF}{{\cal F}}
\newcommand{\cH}{{\cal H}}
\newcommand{\cB}{{\cal B}}
\newcommand{\cO}{{\cal O}}

\documentclass[12pt]{article}
\begin{document}
\renewcommand{\thefootnote}{\fnsymbol{footnote}}
\font\csc=cmcsc10 scaled\magstep1
 {\baselineskip=14pt
  \rightline{
  \vbox{\hbox{UT-907}
        \hbox{August 2000}
 }}}
% \begin{flushright}
% {\sl Preliminary draft}\\
% {\sl \today} 
% \end{flushright}
\vfill
\begin{center}
{\Large
Topological Charges
of Noncommutative Soliton
}\\

\vfill

{\csc Y. Matsuo}\footnote{
      e-mail address : matsuo@hep-th.phys.s.u-tokyo.ac.jp}
\vskip.1in

{\baselineskip=15pt
\vskip.1in
  Department of Physics \\
Faculty of Science,  University of Tokyo\\
%  Bunkyo-ku, Hongo 7-3-1 \\
Tokyo 113-0033, Japan
\vskip.1in
}

\end{center}
\vfill

\begin{abstract}
{%\baselineskip 15pt
The noncommutative soliton is characterized by the use
of the projection operators in non-commutative space.
By using the close relation with the K-theory
of $C^*$-algebra, we consider the variations of 
projection operators along the commutative directions and 
identify their topological charges.
When applied to the string theory, it gives
the modification of the brane charges due to tachyon background.
}
\end{abstract}
\vfill

%\vskip.3in
hep-th/0009002
\setcounter{footnote}{0}
\renewcommand{\thefootnote}{\arabic{footnote}}
\newpage
%%%%%%%%%%%%%%%%%%%%%%%%%%%%%%%%%%%%%%%%%%%%%%%%%%%%%%%%%%%%%%%%%%%%%%
\section{Introduction}
In the recent development of string theory, 
the study of the D-brane has been one of the most fruitful
sources of the inspirations. 
One of the interesting issues among them is to understand
the dynamical process such as brane-anti-brane pair annihilation.
Research toward such direction was pioneered by A. Sen \cite{r:tachyon} 
and the r\^ole of the tachyon field was clarified.
Inspired by these works, E. Witten \cite{r:Witten2} proposed that the 
D-brane charge should be measured by the $K$-group
instead of the cohomology.  It is based on the fact that
D-brane carries the information of the vector bundle due to
the massless gauge fields on it.

A conceptual progress was made by the recent discovery of the 
non-commutative soliton \cite{r:GMS}.
The solution is characterized by the use of  the projection
operators\footnote{
We note that the significance of the projection operator
was already noticed in \cite{r:Fu} in the context
of the noncommutative Yang-Mills theory.
}
which reflects the very nature of the noncommutativity.
The idea was applied immediately \cite{r:Chicago}\cite{r:Indian} 
to the string theory as a mechanism to understand
the tachyon condensation.
The simplification in the large non-commutative
limit helps the analysis by justifying to neglect
many stringy corrections in the lagrangian.

{}From the mathematical viewpoint, the use of projection operators
are quite suggestive since they play a fundamental r\^ole 
to study the geometry of the non-commutative space.
% For example, the index theorem is formulated in
% terms of the cyclic cohomology of the projection operators.
In the context of the $K$-theory of $C^*$-algebra 
(see for example \cite{r:WO}),
the $K_0$ group of a $C^*$-algebra ${{\cal A}}$ is identified as the
equivalence class of the  formal differences 
of the projection operators in 
${{\cal A}}$. In this viewpoint, the use of projection operators
in the tachyon condensation gives a natural passage
from topological to $C^*$-algebraic description of $K$-theory.

In \cite{r:Conne} and the references therein, we can find
the  topological invariants constructed for arbitrary 
$C^*$ algebra.  The simplest example is
the rank $(=Tr(\Pi))$ of the projection operator $\Pi$.
In the context of the tachyon condensation \cite{r:Chicago},
it was identified as the number of 
the D-branes of the lower dimensions 
which are created by the tachyon condensation process
(we call it ``the descendent D-brane'' in the following).
More general invariants can be constructed by
pairing the projection operator with the element $\varphi$
of the cyclic cohomology of the $C^*$-algebra ${{\cal A}}$,
\begin{equation}\label{invariants}
\varphi(\underbrace{\Pi,\cdots,\Pi}_{2n+1})\sim \int \mbox{Tr}\left(\Pi (d\Pi)^{2n}\right)\,\,,
\end{equation}
for positive integers $n$. Here $d$ is the ``derivative''
operator which defines the cyclic cohomology.
Conne used this pairing to define the non-commutative
version of the index theorem. 

One purpose of this note is to identify the
D-brane interpretation of the topological charges
(\ref{invariants}). In the tachyon condensation process,
the relevant $C^*$-algebra is identified as
${{\cal A}}\equiv C^\infty(M)\otimes \cB(\cH)$ where $M$ is the world
volume for the descendent brane and $\cB(\cH)$ is the
linear operators acting on the Hilbert space $\cH$
of the harmonic oscillators. 
It is actually the simplest example as the cohomology
of $C^*$-algebra. In fact, it is well-known \cite{r:WO} that
the $K_0({\cal A})$ is isomorphic to
$K^0(M)$, the {\em topological} $K$  group of the manifold $M$.
In other word, there is a direct correspondence between
the projection operator in ${{\cal A}}$ and
the vector bundle on $M$. We identify the projection
operator as defining noncommutative soliton and
the vector bundle as those of the descendent brane.
In this context, the derivative $d$ in (\ref{invariants})
reduces to the ordinary exterior derivative along $M$ and the
charges (\ref{invariants}) becomes the characteristic
class of the vector bundle on $M$.
This correspondence gives additional
insights, namely the r\^ole of the tachyon background,
to the origin of the gauge symmetry on the descendent D-brane.

\section{Noncommutative soliton and topology}
To describe the idea in more physical language,
we start from the scalar field theory in 
$(2q+2)$ spatial dimensions
($q\geq1$) slightly extending \cite{r:GMS} by
introducing extra commutative directions.
We assume there are two non-commutative 
directions (described by coordinates $y,\bar y$)
and $2q$ commutative directions described by $x^a$.

For the static configuration, the energy functional is given as,
\begin{equation}\label{energy}
 E=\frac{1}{g^2}\int d^{2q} x  d^2 y\, 
(\frac{1}{2}\partial_x\phi\partial_x\phi +
\frac{1}{2}\partial_y\phi\partial_{\bar y}\phi + V(* \phi))\,\,.
\end{equation}
Here $*$-product is defined by a non-commutativity parameter
$\theta$ as
\begin{equation}\label{star-product}
 A * B = \left.e^{\frac{\theta}{2}(\partial_y\partial_{\bar{y}'}-\partial_{y'}
\partial_{\bar{y}})} A(x: y,\bar y) B(x:y',\bar y')
\right|_{y=y',\bar y=\bar y'}
\end{equation}
In large $\theta$ limit, we rescale, $
 y\rightarrow \sqrt{\theta}  y$,
$ \bar y\rightarrow \sqrt{\theta}  \bar y
$.  After the rescale, the first and the third terms
in (\ref{energy}) are multiplied by $\theta$
and in the infinite $\theta$ limit the second term 
can be neglected.

% to obtain the modified energy functional,
% \begin{equation}
%  E=\frac{1}{g^2}\int d^{2q} x  d^2 y\, 
% (\frac{\theta}{2}\partial_x\phi\partial_x\phi +
% \frac{1}{2}\partial_y\phi\partial_{\bar y}\phi + \theta V(* \phi))\,\,,
% \end{equation}
% where $*$-product (\ref{star-product}) with $\theta=1$ is used
% in the potential term. In large $\theta$ limit, one can neglect 
% the second term (derivative in $y$ directions) in the energy functional.

GMS soliton is constructed by the
field configuration satisfying
\begin{equation}\label{projection}
 \phi_0(y) * \phi_0(y) = \phi_0(y)\,\,.
\end{equation}
Indeed if $\lambda_*$ gives the minimum of the potential
$\left.\frac{\partial V(\lambda)}{\partial \lambda}\right|_{\lambda
=\lambda_*}=0$, \newline
$\phi(y)=\lambda_* \phi_0(y)$ also minimizes the
potential,
% \begin{equation}
%  \left.\frac{\partial V(\phi)}{\partial \phi}\right|_{\phi=\lambda_* \phi_0}
%  = \left. \frac{\partial V(\lambda)}{\partial \lambda}
% \right|_{\lambda=\lambda_*}\phi_0(y) = 0\,\,,
% \end{equation}
as long as $V(\lambda)$ is a polynomial of $\lambda$.

For the explicit construction of the configuration
which satisfies (\ref{projection}), GMS used one-to-one
correspondence between the space of the functions of 
the non-commutative coordinates (say $\cF$) and the space of
the linear operators $\cB(\cH)$ acting on the Hilbert space $\cH$
of the harmonic oscillators. The correspondence uses the 
Weyl ordering and is defined as,
\begin{eqnarray}
 f(y,\bar y)\in \cF & \leftrightarrow & \cO_f\in \cB(\cH)\nonumber\\
 \cO_f & = & \frac{1}{2\pi} \int d^2 k \tilde{f}(k,\bar k)
 e^{2\pi i (k\bar a^\dagger + \bar k a)}\\
 \tilde{f}(k,\bar k) & \equiv & \int d^2 y f(y,\bar y) e^{-2\pi i 
(y\bar k + k\bar y)}\,\,.\nonumber
\end{eqnarray}
The Moyal product in $\cF$ is translated
into the ordinary product in $\cB(\cH)$,
\begin{equation}
 \cO_f\cdot \cO_g = \cO_{f*g}\,\,.
\end{equation}
Therefore in $\cB(\cH)$, (\ref{projection}) is translated into the
conventional condition of the projections,
$\Pi^2=\Pi$  for $\Pi = \cO_{\phi_0}$.

In $\cB(\cH)$ the construction of the projections is straightforward
since we know the orthonormal basis $|k\rangle=\frac{{a^{\dagger}}^k}{
\sqrt{k!}}|0\rangle$. 
% the projection to this basis is
% \begin{equation}
% |k\rangle \langle k| = \frac{1}{k!}:{a^\dagger}^{k}
% e^{-a^\dagger a}a^k: .
% \end{equation}
% Here $::$ is the normal ordering. 
% s shown by GMS
% that wave function associated with this projector
% is written with $k$-th Laguerre polynomial.
GMS introduced the  rank $k$ solution as the wave function
corresponding to,
\begin{equation}
 \Pi_k \equiv \sum_{r=0}^{k-1} |r\rangle \langle r|\,\,.
\end{equation}
By putting it into the potential, the energy
can be evaluated  as 
\begin{equation}
 \int d^2 y V(\lambda_* \phi_k(y,\bar y) ) =\mbox{Tr}_\cH
 (V(\lambda_*) \Pi_k) = V(\lambda_*) k\,\,,
\end{equation}
where we denote $\phi_k(y,\bar y)$
as the wave function corresponding to $\Pi_k$.

Actually there are infinite number of projection operators
which have the same potential energy.  They are obtained by twisting
projection operators by $g\in U(\cH)$ where $U(\cH)$ is the 
set of unitary operators acting on $\cH$,\footnote{
GMS has shown that the kinetic term
favors the choice $g=1$ in finite $\theta$.
In our context, it will help to give a unique
solution to each topologically disconnected
sectors of projection operators.
}
\begin{equation}
 \Pi^g_k=g\Pi_k g^{-1}\,\,.
\end{equation}
The set of rank $k$ projection is thus parametrized by the
infinite dimensional grassmannian 
\begin{equation}
Gr_{k}(\cH)\equiv\left\{
\mbox{set of $k$ dimensional subspaces in }  \cH
\right\}\,\,,
\end{equation}
since $\Pi^g_k$ is specified by the $k$-dimensional subspace
to which projection is defined.
These spaces have rich topology. 
For example when $k=1$ it is
simplified to the infinite dimensional projective space $CP(\cH)$
which has the homotopy group, $\pi_{2q}(CP(\cH))={\bf Z}$
for $q=0,1,2,3\cdots$.

In the presence of the extra commuting space $M$,
one may consider the variations of the
projection operators along that direction.
This is, of course, valid only if the potential is 
dominant enough compared to the energy coming from
the variation in $x$ directions.
For that purpose, one may rewrite $V(\phi)$ as $t V(\phi)$
and takes $t\rightarrow \infty$ limit.
%  while adjusting
% the minimum of the potential to zero $V(\lambda_*)=0$.
In this limit, one may restrict the configuration
space of $\phi$ to the space of rank $k$ projections in $\cH$
and the configuration space of the scalar field theory in 
$(2q+2)+1$ dimensions is reduced to the non-linear
sigma model in $2q+1$ dimensions whose target space is $Gr_k(\cH)$.
This is physically interpreted as the Nambu-Goldstone bosons 
associated with the symmetry breaking 
$U(\infty)\rightarrow U(k)\times U(\infty-k)$.

{}From the mathematical viewpoint,
they give the general idempotent elements
in $C^\infty(M)\otimes \cB(\cH)$
and should be included to the analysis of the $K$-theory
of $C^*$-algebra.
% We note that such operators
% in general include the dependence on the coordinates of $M$
% since there is no a priori reason to 
% restrict it to the constant one.
As we wrote in the introduction, there is a correspondence
between the projection operator of $C^\infty(M)\otimes \cB(\cH)$
and the vector bundle over $M$. From the projection operator
$\Pi$, we define the fiber bundle on $M$ by
assigning a fiber $\Pi_x(\cH)$ for each point $x\in M$.

Actually $Gr_k(\cH)$ is the {\em classifying space}
of rank $k$ vector bundles.  Namely
any vector bundle over $M$ is isomorphic to
the vector bundle thus defined by the projection
operator. The homotopy class of the vector bundle
is classified by the mapping
\begin{equation}
 M\rightarrow Gr_k(\cH)\,\,.
\end{equation}
In this sense, there is a one-to-one correspondence between
the homotopy class of the vector bundle over $M$ and
the connected components of the configuration
space of the non-linear sigma model.

The remaining task is to identify (\ref{invariants})
with the characteristic class of the vector bundle
by using the projection operators.
This is an elementary material but let us write down
explicitly. We need to introduce define the covariant derivative.
by using the projections.

Since the fiber bundle
$\cH\rightarrow M$ itself is trivial and just the direct product,
all the topological non-trivialities come from the projection
onto the finite dimensional subspace.
The covariant derivative in such a situation is defined as,
\begin{equation}
 D \equiv \Pi(x) \cdot d\cdot \Pi(x).
\end{equation}
To relate it to the conventional definition
of the covariant derivative, we introduce the coordinate
dependent orthonormal basis,
\begin{equation}
 \langle i |j \rangle =\delta_{ij},\quad
 \Pi(x) =\sum_{i=0}^{k-1} |i \rangle \langle i|\,\,.
\end{equation}
The sections of the vector bundle is written locally 
as $f(x)=\sum_{i=0}^{k-1} f_i(x)|i\rangle$. 
The covariant derivative acts on it as
\begin{equation}
 D f(x) = \sum_{i=0}^{k-1} (Df)_i |i\rangle\,\,,\quad
 Df_i = df_i +\sum_{j=0}^{k-1}  A_{ij} f_j\,\quad
 A_{ij} = \langle i | d | j\rangle\,\,.
\end{equation}
$A_{ij}$ gives the  $U(k)$ gauge connection
which is an analogue of the  non-abelian Berry phase \cite{r:Berry}.
The local change of the basis $|i\rangle$ gives the
gauge transformation.
This gauge symmetry is originated from
the configuration space of projection operators
namely the Grassmaniann $Gr_k(\cH)$.
It is well-known that there is a ``hidden''
gauge symmetry of gauge group $H$ for the 
coset space nonlinear sigma model on $G/H$.

% The derivative of this section does  not in general
% become a section of vector bundle.  To recover it
% as a section, we need to apply the projection operator again.
% This is the definition of the covariant derivative and summarized as,
% \begin{equation}
%  D \equiv \Pi(x) \cdot d\cdot \Pi(x) = d + A\,\,.
% \end{equation}
% $A$ takes its value in Lie algebra of $U(k)$.
% Locally the projection operator can be expressed in terms of
% the orthonormal frame $\left\{e_i(x)\right\}$ $(i=0,1,2,\cdots)$
% satisfying $e_i(x)\cdot e_j(x)=\delta_{ij}$ as 
% $\Pi=\sum_{i=0}^k{}^te_i(x) e_i(x)$.
%In this frame, the $U(k)$ connection is expressed as
%$A_{\mu i j}=e_i\cdot\partial_\mu e_j$.

The curvature two form associated with this covariant derivative
is,
\begin{equation}
 F=D^2 = \Pi d\Pi d \Pi\,\,.
\end{equation}
{}From this expression, it is straightforward to write
the characteristic class associated with the vector 
bundle defined by the projection.
For example, $p$-th Chern class is given by (\ref{invariants}),
$ c_p\propto \int \mbox{Tr}\, F^p  = \int \mbox{Tr}\, \Pi\,(d\Pi)^{2p}$.

\section{Application to string theory}
% The physical interpretation of these topological charges is
% clearer when applied to the string theory.
In the string theory, the commuting
directions are identified with the world volume
of the descendent D-brane.
Let us first consider the open bosonic string 
which has boundaries at one D-$(p+2)$ brane \cite{r:Chicago}.
In this context, the scalar field is replaced by
the tachyon field $T(x,y)$ and there is also 
the noncommutative $U(1)$ gauge field
$A_\mu(x,y)$ on the brane. The action is
\begin{equation}
 S\sim \int dt d^{p} x d^2 y\left(
f(* T) (D T)^2 -V(* T) + g(* T)(F_{\mu\nu})^2
+\cdots\right).
\end{equation}
%
% $*$-product is induced by the constant $B_{\mu\nu}$ field 
% in $y$ directions. 
% $\cdots$ is the stringy corrections which should be
% higher order in $\alpha'$. 
% $D$ is the covariant derivative
% associated with the non-commutative $U(1)$ gauge symmetry
% described by $A_\mu$. 
% and the limit where extra terms can be neglected, we refer
% the paper \cite{r:Chicago}.
%
The functions $f(T), V(T), g(T)$ can be in principle
determined by the string field theory \cite{r:SFT}.
As conjectured by Sen \cite{r:Sen}, the information
to construct non-commutative soliton is supplied by the
``universality'' of these functions.  It is conjectured
that there are two critical points $T=0, t_*$ in $V(T)$.
When $T=t_*$, $V(t_*)$ is equal to the tension of the
D-branes.  This point is the ordinary perturbative
vacuum of the open string.
% At this point tachyon energy should cancel the 
% D-brane tension since D-brane can  decay to the vacuum
% where we have no energy.
When $T=0$, we have $f(0)=g(0)=V(0)=0$.
This is the point where the D-brane is annihilated to vacuum 
and there are no propagating open string degree of freedom.
For the recent development of the ``nothing state'', see \cite{r:nothing}.

The noncommutative soliton solution for the tachyon field 
is given by,
\begin{equation}
 T(x,y) = t_* \cdot \phi(y) + 0\cdot (1-\phi(y)).
\end{equation}
Here $\phi(y)$ is the configuration corresponding to
degree $k$ projection $\Pi_k$.  
Following to Witten's notation \cite{r:Witten},
we denote $V$ as the subspace in $\cH$ which is defined by the
projection $\Pi_k$ and $W$ is specified by $1-\Pi_k$.  
% When $\Pi$ varies along the world volume $M$, $V$ and $W$ defines
% fiber bundles over $M$.
In \cite{r:Chicago}, it is shown that
the soliton solution can be naturally identified as $k$
D-$p$ branes which emerge after the tachyon condensation.

Around this vacuum, the fluctuation of
tachyon fields and the gauge fields can be expanded
in terms of $\phi_{nm}(y)$ which corresponds to
$|n\rangle \langle m|$ in $\cB(\cH)$.
% By applying the projection operators $\Pi$ and $1-\Pi$,
% these are classified into four sectors,
% $VV$, $VW$, $WV$, and $WW$. 
In particular, $U(1)$-connection $A_\mu$ in $p+2$-brane
can be identified as $U(\cH)$-connection on $p$-brane.
By sandwiching it by the projection operators
$\Pi$ and $1-\Pi$, we get four sectors.
Namely for $\cO\in \cB(\cH)$, we write
\begin{eqnarray}
\cO_{VV}\equiv \Pi\cO\Pi\,\,, &\quad&
\cO_{VW}\equiv \Pi\cO(1-\Pi)\,\,,\nonumber\\
\cO_{WV}\equiv (1-\Pi)\cO\Pi\,\,,&\quad&
\cO_{WW}\equiv (1-\Pi)\cO(1-\Pi)\,\,.
\end{eqnarray}
Components in $VW$, $WV$, and $WW$ sectors become non-propagating
after the tachyon condensation since they are connected
with the nothing states. 
% In \cite{r:Chicago}, it is explained that
% the gauge symmetry on the descendent brane comes from the 
% $VV$ projection of $U(\cH)$ gauge  fields.
The $VV$ sector, on the other hand, describes the surviving
physical modes in the descendent D-branes.

At this point we introduce the position dependent projection operators.
As in the field theory example, it induces the
gauge symmetry on the world volume of the descendent brane.
Since we already have such gauge symmetry from $VV$ part
of the $U(\cH)$ gauge field $A$ \cite{r:Chicago}, the position
dependent projection operators introduce the 
modifications of the covariant derivative,
\begin{equation}
 \Pi (d+A) \Pi \equiv \Pi D_A \Pi= \Pi\,d\,\Pi + A_{VV}\,\,,
\quad
A_{VV}\equiv \Pi A \Pi.
\end{equation}
{}From this expression, the effective curvature is obtained as,
\begin{equation}\label{modified}
 \cF = \Pi (D_A \Pi)^2=F_{VV} + \Pi(d\Pi)^2+\cdots\,\,.
\end{equation}
Here $F_{VV}=dA_{VV}+A_{VV}\wedge A_{VV}$ is the curvature form
from $A_{VV}$. 
% This effective curvature defines the
% structure of the vector bundle over the $p$-brane.
The second term is the correction of the curvature
from the tachyon configuration.

We may understand the implication of this formula
in the following way. We have two equivalent routes
to realize a nontrivial vector bundle over the 
descendent D-brane created by the tachyon condensation.  
One track is to start
from the nontrivial $U(\cH)$ bundle and apply the constant projection.
The other is to apply the twisted projection to the trivial 
$A=0$ background.
% In the second approach, the information of the vector bundle
% is directly connected with the idempotent elements
% in noncommutative algebra ${{\cal A}}$.
The first viewpoint has a merit to understand
the relation between gauge fields in the original
D-brane and the descendant.
On the other hand the second approach is better
to understand the direct relation with the 
K-theory of the $C^*$-algebra.
% $C^\infty(M)\otimes \cB(\cH)$.

% It is well-known that the arbitrary vector bundle ($P$) on the manifold
% $M$ can be embedded into the trivial vector bundle $M\times {\bf C}^n$
% if $n$ is sufficiently large. From the trivial vector
% bundle, $P$ is defined by the $x$ dependent projection 
% operators on the fiber $\cH$.  In our case, it means that
% arbitrary gauge bundle over $M$ can be constructed
% even if we have vanishing gauge field $A=0$
% before the tachyon condensation.

At this point, it is pedagogical to
indicate an analogy with the 't Hooft-Polyakov monopoles 
\cite{r:monopole} where Higgs field varies along the spatial infinity
and so is the projection to the unbroken $U(1)$ part.
For the calculation of the monopole charge, one needs to modify
the $U(1)$ field strength by the Higgs field.  For example,
in the simplest $SU(2)\rightarrow U(1)$ case, the effective
field strength is given as,
\begin{equation}
 \cF_{\mu\nu} = F_{n\mu\nu} \hat\phi_n - \epsilon_{nml}
\hat\phi_n D_\mu \hat\phi_m D_\nu \hat\phi_l\,\,.
\end{equation}
Here the Higgs field ($\hat\phi_n\equiv
\phi_n/\sqrt{\phi^2}$) takes their value in $S^2$
where the Higgs potential is minimized. 
% The analogy 
% between this formula and (\ref{modified}) is manifest.
The inclusion of the latter term is essential to
evaluate the monopole charge as the
winding number of $\pi_2(S^2)$.

The correspondence between $K$-theory of
$C^*$-algebra and tachyon condensation becomes more accurate
in the pair annihilation of $D$-$\bar{D}$ 
system in the superstring.
In commutative case, we already know
\cite{r:Witten2} that the formal difference between the 
vector bundles defined on $D$ and $\bar{D}$ branes defines
the topological K-group, i. e. the D-brane charges in $M$.
In the noncommutative case, similar definition of
$K$-group is possible as the formal difference
between the  two projection operators.
Such an idea was proposed in \cite{r:Witten}.
We start from  a pair of D-$(p+2)$ brane and
$\bar D$-$(p+2)$ brane and introduce the large
non-commutativity in two directions.
% the $U(1)$ gauge field
% on the D-$(p+2)$ brane (resp. $\bar D$-$(p+2)$ brane)
% is expanded as to define $U(\infty)$ connection
% on the D-$p$ brane (resp. $\bar D$-$p$ brane).
In this case, we have two copies of
the same Hilbert space of the harmonic oscillators 
on $D$- (resp$\bar D$-) brane. We denote them as $\cH$ and $\bar\cH$.
The tachyon fields $\sigma,\bar\sigma$ 
appear as the {\em interpolating} 
linear map between the two Hilbert spaces,
\begin{equation}
 \sigma : \cH \rightarrow \bar\cH\,\,\quad
 \bar\sigma : \bar\cH \rightarrow \cH\,\,,
\end{equation}
with the finite dimensional kernel and cokernel
(i.e. Fredholm operators).
We note that the tachyon fields themselves can not
be idempotent as in the bosonic string. Witten \cite{r:Witten}
has shown that they should instead satisfy
\begin{equation}
 \sigma \bar\sigma \sigma=\sigma\,\qquad
 \bar\sigma \sigma \bar\sigma = \bar\sigma\,\,,
\end{equation}
to recover the equation of motion of the 
string field theory.  
% These relations are indeed natural
% objects in the $K$-theory of the ${\bf C}^*$-algebra
% (for example see \cite{r:WO}) where $\sigma$ is called partial isometry. 
{}From such operators, one may construct two projection operators
acting on $\cH$ and $\bar\cH$,
\begin{equation}
 \Pi=1-\bar\sigma\sigma\,\,,\quad
 \bar\Pi=1-\sigma\bar\sigma\,\,.
\end{equation}
It is easy to observe that $\Pi$ (resp. $\bar\Pi$) defines 
the projection to the kernel $V$ (resp. cokernel $W$) of 
$\sigma$ in $\cH$ ($\bar\cH$)
\begin{equation}
 \Pi^2=\Pi,\quad
 \bar\Pi^2=\bar\Pi,\quad
 \Pi\bar\sigma = \bar\Pi\sigma =0\,\,.
\end{equation}
The dimensions of $V$ and
$W$ are identified as the number of 
descendant $D$-$p$ ($\bar D$-$p$)
branes after the tachyon condensation. The index of $\sigma$,
\begin{equation}\label{index}
\mbox{Ind}(\sigma)=\mbox{dim}(\mbox{Ker }\sigma)
-\mbox{dim}(\mbox{Coker }\sigma)=
\mbox{Tr}_{\cH}\,\Pi -\mbox{Tr}_{\bar\cH}\,\bar\Pi\,\,.
\end{equation}
is the total $D$-$p$ brane number.

We repeat the generalization where the partial isometry $\sigma$
varies along the world volume.
As in the case of the bosonic string,
$\Pi$ (resp. $\bar\Pi$) defines a vector bundle on
$D$- (resp. $\bar D$-) $p$-brane world volume $M$. 
The left hand side of (\ref{index}) 
becomes the index bundle, namely the formal difference
between two vector bundles $[\mbox{Ker}(\sigma)]-
[\mbox{Coker}(\sigma)]$ over $M$.
The isomorphic class of the index bundle precisely
defines an element of the topological $K^0(M)$ group.

In the presence of the non-constant tachyon field $\sigma$, 
the D-brane charges embedded in the descendant 
brane should be also modified.
The result is completely parallel to
the bosonic case. The field strengths
on $D$ and $\bar{D}$ branes should be modified to
\begin{equation}
 \cF=\Pi(D_A \Pi)^2\,\,,\quad
 \bar\cF=\bar\Pi(D_{\bar A} \bar\Pi)^2\,\,.
\end{equation}
Such a change forces us to modify the
D-brane charges on the world volume.
They are evaluated from
Chern-Simons coupling\cite{r:CS} with modified
gauge potential,
\begin{equation}\label{brane_charge}
  \int_{M} C \wedge
  (\mbox{Tr}_\cH(e^\cF)-\mbox{Tr}_{\bar\cH}(e^{\bar\cF})).
\end{equation}
We note that it will be useful if this formula 
can be ``derived'' from the generic result for
the commutative case derived by
\cite{r:KW} who used the concept of the superconnection \cite{r:Q},
\begin{equation}
 \int C\wedge \mbox{STr} e^{\cF}\,\,,\quad
 \cF=\left(
\begin{array}{c c}
 F_{VV}-T\bar T& DT\\
 \overline{DT} & F_{WW} -\bar T T\,\,
\end{array}
\right)\,\,.
\end{equation}

\section{Future directions}
There are a few directions which we would like to
proceed in the future.  One direction is the generalization
of the $C^*$-algebra and the cyclic cohomology. 
Originally Conne introduced such machinery
to study the geometrical objects which is not
accessible from the topological methods.  There are a lot
of examples which seem to be described in the context of
noncommutative solitons.  The second direction is
the use of $K_1({{\cal A}})$ group.  In the context of $C^*$ algebra,
they are defined by the unitary operators on the algebra.
In the commutative context, Horava \cite{r:Horava} argued that
tachyon field defines the unitary transformation that
defines the vector bundle on the brane. 
It would be critically important to find the analogue of
noncommutative solitons constructed out of the unitaries.
For the current understanding to this direction, 
we can find many interesting suggestions in \cite{r:Witten3}.

\vskip 10mm
Note added: After we submitted this paper on the network, 
we realized that Harvey and Moore \cite{r:HM} were 
independently working on an idea which is very close to ours.

\vskip 10mm
\noindent{\sl Acknowledgements:}\hskip 2mm
{\em \small The author would like to thank the organizers of the workshop
``Summer Institute 2000 at Yamanashi'' where he could
enjoy comfortable environment to discuss
with other participants and have started this work.
He is also  obliged to S. Terashima and T. Takayanagi
for valuable discussions on noncommutative solitons
and especially E. Ogasa for explaining the 
mathematical background. 

The author is supported in part by Grant-in-Aid 
(\# 09640352) and in part by Grant-in-Aid for Scientific
Research in a Priority Area, ``Supersymmetry and Unified
Theory of Elementary Particle'' (\# 707) from
the Ministry of Education, Science, Sports and Culture.}

%%%%%%%%%%%%%%%%%%%%%%%%%%%%%%%%%%%%%%%%%%%%%%%%%%%%%%%%%%%%%%
%\newpage

\end{document}